\documentstyle[pra,aps,multicol,epsf]{revtex}
\input psfig
\def\be{\begin{equation}}
\def\ee{\end{equation}}

\begin{document}
\bibliographystyle{simpl1}

\title{Exact Eigenstates for Repulsive Bosons in Two Dimensions}

\author{R A Smith and N K Wilkin}
\address{School of Physics and Astronomy, University of
Birmingham, Edgbaston, Birmingham B15 2TT, ENGLAND}
\maketitle
\bigskip

\begin{abstract}
We consider a model of  $N$ two-dimensional bosons in a harmonic
potential with weak repulsive delta-function interactions.
We show analytically that, for angular momentum $L\le N$, the elementary
symmetric polynomials of particle coordinates measured from the center of
mass are exact eigenstates with energy $N(N-L/2-1)$. Extensive numerical
analysis confirms that these are actually the ground states, but we are
currently unable to prove this analytically. The special case $L=N$ can be
thought of as the generalisation of the usual superfluid one-vortex state
to Bose-Einstein condensates in a trap.
\end{abstract}

\begin{multicols}{2}

\section{Introduction}

The recent observation of Bose-Einstein condensation in dilute gases
of alkali metals \cite{And95,Dav95,Brad97} has stimulated much interest 
in the properties of
systems of interacting bosons. A question of particular interest is
whether such systems will form vortices under rotation, as occurs in
superfluid $^4$He. Experimentally such vortex states have been observed,
both in two component systems \cite{Mat99} and in a stirred condensate of 
$^{87}$Rb atoms \cite{Mad00}. Theoretically the stability of such vortices
has been considered both in the Thomas-Fermi limit of strong interactions
between atoms using the Gross-Pitaevskii equation \cite{Rok97,Butt99}, 
and also in the weak interaction limit 
\cite{Wil98,Coop99,Wil00,Mot99,Ber99,Kav00,Jack00} 
where the coherence length is much larger than the size
of the atom cloud. It is the latter limit we shall focus upon in this paper.

Pursuing an analogy with the fractional quantum Hall effect \cite{Trug85},
we previously introduced a model of weakly interacting bosons in
a harmonic well \cite{Wil98} to address the question of whether attractive 
bosons condense. We proved analytically that all the angular momentum 
in this model is carried by the center of mass for attractive bosons, 
whereas for repulsive bosons we numerically found that a vortex
state forms in the case of one unit of angular momentum per boson. 
Further numerical work by one of the authors \cite{Coop99,Wil00} 
extensively investigated the
properties of ground states of the repulsive model for angular momentum
$L>N$, demonstrating
that although such states are more complicated than the analytic ones known
for $L\le N$, they can still be understood either within vortex or composite
fermion or boson pictures. Mottelson \cite{Mot99} considered the low-lying 
eigenstates 
for the case $L\le N$, whilst Bertsch and Papenbrock \cite{Ber99} performed
numerical computations and noted that the ground state of the repulsive model
for $L\le N$ is the elementary symmetric polynomial, $\tilde{e}_L$, of coordinates, 
$z_i=x_i+iy_i$, relative to the center of mass, $R=\sum_i z_i/N$.
Finally, recent work by Kavoulakis et al \cite{Kav00} and Jackson et al \cite{Jack00}
has considered the relationship between mean and exact numerical solutions
in the limit of large $N$.

In this Letter we present an analytical proof that the state described
above is an exact eigenstate of the model for $L\le N$. We have unfortunately
been unable to show that this state is the ground state, although we have
considerable numerical evidence to show that this is the case.  

The model is of $N$ bosons in a harmonic potential in two dimensions
interacting via a delta--function potential, for which the Hamiltonian is
\begin{equation}
\label{Ham}
H=\sum_{i=1}^{N} \left[-{\hbar^2\over 2m}\nabla_i^2
+{1\over 2}m\omega^2r_i^2\right]
+V\sum_{i<j}\delta(\mathbf{r}_i-\mathbf{r}_j)
\end{equation}
The natural way to look at this problem is in second quantised form, so
first we must solve the non-interacting problem. We do this in plane polar
coordinates since we are interested in angular momentum properties. The
one-particle wavefunctions and energies are then
\begin{eqnarray}
\label{2DSHO}
\psi_{n_r,\ell}&=&N_{n_r,\ell}R(n_r,|\ell|,-r^2/2)r^{\ell}e^{i\ell\theta}
\nonumber\\
E_{n_r,\ell}&=&(n_r+|\ell|+1)\hbar\omega
\end{eqnarray}
where $R(n,l,x)$ is the confluent hypergeometric function, $n_r$ is the radial
quantum number and $\ell$ is the angular momentum. The total energy and
angular momentum of a system of $N$ non-interacting bosons in this harmonic
well is thus
\begin{equation}
\label{N2DSHO}
E_{tot}=\sum_{i=1}^{N} (n_{r,i}+|\ell_i|+1)\hbar\omega,\qquad
L_{tot}=\sum_{i=1}^{N} \ell_i.
\end{equation}
The ground state manifold is then obtained by putting all bosons into the
lowest radial state, $n_{r,i}=0$, and choosing angular momentum $\ell_i$
all of the same sign (which we choose to be positive), such that
\begin{equation}
\label{GSman}
\sum_{i=1}^{N} \ell_i = L\quad\Rightarrow\quad
E_{tot}=(L+N)\hbar\omega.
\end{equation}
It can be seen that the ground state manifold has a degeneracy which is
equal to the number of ${\it partitions}$ of the total angular momentum
$L$ into the $N$ angular momenta $\ell_i$ of the individual bosons.

If we assume that the dimensionless interaction strength is very small,
$|\eta|\ll 1$, where $\eta=V/\hbar\omega$, we can treat it as a 
perturbation whose sole effect will
be to lift the large degeneracy of the non-interacting ground state. This
means we can work within the non-interacting ground state manifold, which
freezes out the one-particle part of the Hamiltonian, and consider the
effect of the interaction using degenerate perturbation theory. This is
essentially an extension of an approach used for fractional quantum Hall
effect systems \cite{Trug85} to bosons. The normalised one-particle wavefunctions are
given by
\begin{equation} 
\label{Wfns}
\psi_{\ell}(r,\theta)=
{1\over\sqrt{\ell!2\pi}}r^{\ell}e^{i\ell\theta}e^{-r^2/2}
={1\over\sqrt{\ell!2\pi}}z^{\ell}e^{-|z|^2/2},
\end{equation}
where we have moved to complex notation. The second quantized form of the
interaction Hamiltonian is thus
\begin{equation}
\label{2QHam}
\hat{H}=\sum_{m_1,m_2,n_1,n_2} V_{m_1m_2n_1n_2}
a_{m_1}^+a_{m_2}^+a_{n_1}a_{n_2}
\end{equation}
where the matrix element
\begin{equation}
\label{Vint}
V_{m_1m_2n_1n_2}
={\eta\over 4\pi}{(m_1+n_1)!\over 2^{m_1+m_2}}\delta_{m_1+m_2,n_1+n_2}.
\end{equation}
For future convenience we will set $\eta=4\pi$.

To perform the degenerate perturbation theory for given $N$ and $L$,
we first need the basis states which are labelled by the partitions of
$L$ into $N$ pieces. Let us write a partition $\lambda$ in the form
\begin{equation}
\label{parts}
\lambda=0^{\lambda_0}1^{\lambda_1}2^{\lambda_2}\dots
=\prod_{i} i^{\lambda_i}
\end{equation}
where
$\sum\lambda_i=N$
and
$\sum i\lambda_i=L$.
The corresponding basis state is then
\begin{equation}
\label{basis}
|\,\lambda\,\rangle=
\left[\prod_i {(a_i^+)^{\lambda_i}\over\sqrt{\lambda_i!}}\right],
|\,0\,\rangle
\end{equation}
where $a_{\ell}^{+}$ creates a boson of angular momentum $\ell$.
In coordinate space, this basis state takes the form
\begin{equation}
\label{normwfn}
\left[{\prod_i\lambda_i!\over (2\pi)^{N}N!\,
\prod_i(i!\,)^{\lambda_i}}\right]^{1/2}
m_{\lambda}(z_1,z_2\dots z_N),
\end{equation}
where $m_{\lambda}$ is the {\it monomial symmetric polynomial} corresponding
to the partition $\lambda$. The latter is the symmetric polynomial of the
$N$ variables $(z_1,z_2\dots z_N)$ which has $\lambda_i$ $i$-th powers.

To consider the problem either analytically or numerically requires us to
calculate the elements of the symmetric interaction matrix,
$H_{\lambda\mu}=\langle\lambda|\hat{V}|\mu\rangle$. 
This is obviously best performed using
the second quantized approach. There are two types of matrix elements,
diagonal and off-diagonal, and their evaluation is different. For the
diagonal elements $H_{\lambda\lambda}$, we must sum over every possible
pair of elements in the partition, whether distinct or identical,
\begin{equation}
\label{matrixels}
H_{\lambda\lambda}=\sum_i \lambda_i(\lambda_i-1)
V_{\lambda_i\lambda_i\lambda_i\lambda_i}+
4\sum_{i<j} \lambda_i\lambda_j
V_{\lambda_i\lambda_j\lambda_i\lambda_j}.
\end{equation}
The off-diagonal elements $H_{\lambda\mu}$ can only be non-zero if
$\lambda$ and $\mu$ differ only in the angular momenta of two particles.
This can affect either 4 separate angular momentum states, as in the
case where the angular momentum transfer is $0+4 \rightarrow 1+3$; 
or 3 separate angular momentum states, as in the case of angular momentum transfer
$0+4 \rightarrow 2+2$. In terms of partitions, in case (i) $\lambda_i=\mu_i$ except
at four values of $i$, which we call $i_1\dots i_4$, and
$\lambda_{i_1}=\mu_{i_1}+1$,
$\lambda_{i_2}=\mu_{i_2}+1$, $\lambda_{i_3}=\mu_{i_3}-1$,
$\lambda_{i_4}=\mu_{i_4}-1$. The matrix element is then
\begin{equation}
\label{alldiff}
H_{\lambda\mu}=H_{\mu\lambda}=
4\sqrt{\lambda_{i_1}\lambda_{i_2}\mu_{i_3}\mu_{i_4}}
V_{i_1i_2i_3i_4}.
\end{equation}
In case (ii) $\lambda_i=\mu_i$ except at three values of $i$ which we call
$i_1,i_2,i_3$, and $\lambda_{i_1}=\mu_{i_1}+2$, $\lambda_{i_2}=\mu_{i_2}-1$,
$\lambda_{i_3}=\mu_{i_3}-1$. The matrix element is then
\begin{equation}
\label{twosame}
H_{\lambda\mu}=H_{\mu\lambda}=
2\sqrt{\lambda_{i_1}(\lambda_{i_1}-1)\mu_{i_2}\mu_{i_3}}
V_{i_1i_1i_2i_3}.
\end{equation}
These formulas allow one to write down the interaction matrix $H_{\lambda\mu}$
which can then be diagonalised to give the energy eigenvalues and
eigenstates.

\section{The Subspace Property}

If we look at the original Hamiltonian, we find that the interaction
term depends only upon relative coordinates. To see this, change variables
to the center of mass variable, $R=\sum_i z_i/N$, 
and $N-1$ relative coordinates
such as $\tilde{z}_i=z_i-R$, where $i=1\dots N-1$: the interaction is then a
function only of the relative coordinates $\tilde{z}_i$. 
It follows that if $\psi(z_1\dots z_N)$ is an eigenfunction of $H$ with
a certain energy $E$ and angular momentum $L$, then $R\psi(z_1\dots z_N)$
is an eigenfunction of $H$ with energy $E$ and angular momentum
$L+1$. This means that all eigenfunctions of $H$ at a given $L$ are
reproduced at all higher $L$, and thus the number of new states at any $L$
is just $n_L-n_{L-1}$, the difference between the number of partitions of
$L$ and $L-1$ respectively \cite{Trug85}.

This subspace property makes it natural to think in terms of a second type
of symmetric polynomial, the {\it elementary symmetric polynomials}. These
are defined by, for $N$ variables,
\begin{equation}
\label{elem}
e_L=\sum_{i_1<i_2<\dots<i_L} z_{i_1}z_{i_2}\dots z_{i_L}
\end{equation}
where $L\le N$. For a general partition $\lambda$ we define
\begin{equation}
\label{elemgen}
e_{\lambda}=\prod_i e_i^{\lambda_i}.
\end{equation}
The set of new eigenstates at any total angular momentum $L$ is seen
to be spanned by the $\tilde{e}_{\lambda}$, the elementary symmetric
polynomials of the relative coordinates $\tilde{z}_i$, where we now include
$\tilde{z}_N=-\tilde{z}_1-\tilde{z}_2\dots\tilde{z}_{N-1}$. 
Since $\tilde{e}_1=0$, only partitions
with $\lambda_1=0$ can be formed (i.e. partitions with no part equal
to $1$), which gives exactly the correct number of states.

\section{Proof that the $\mathbf\tilde{\hbox{e}}_L$ are Exact Eigenstates}

In this section we prove that the states $\tilde{e}_L$ for $L\le N$ ($L\ne 1$)
are eigenstates of $\hat{H}$ with eigenvalue $N(N-1-L/2)$. The proof is
essentially a brute force method: we operate the Hamiltonian on the state
$\tilde{e}_L$, and show that the result is the above eigenvalue times
$\tilde{e}_L$. The derivation naturally falls into five steps, which we
detail below. 
\hfil\break
\vskip 0.05truein
\noindent
\noindent{\bf (1) Writing $\mathbf\tilde{e}_L$ in Terms of $\mathbf e_L$ and 
$\mathbf R$}
\vskip 0.05truein
Consider $\tilde{e}_M$, which can be written as
\begin{equation}
\tilde{e}_M=\sum_{i_1<i_2<\dots<i_M} (z_{i_1}-R)\dots(z_{i_M}-R).
\end{equation}
If we expand out this product we will get the elementary symmetric
polynomials of the $z_i$, namely the $e_L$, $L\le M$, multiplied by
$R^{M-L}$. To get the correct coefficients in this expansion we note
that $\tilde{e}_M$ has $N!\,/M!\,(N-M)!$ terms 
whilst $e_L$ has $N!\,/L!\,(N-L)!$ terms.
In the expansion of $\tilde{e}_M$, each product term will produce $M!\,/L!\,(M-L)!$
terms of the type which will add up to produce $R^{M-L}e_L$, so that the
coefficient of $e_L$ in the expansion is ${(N-L)!\,/(N-M)!\,(M-L)!}$ It follows that
\begin{eqnarray}
\label{emexpand}
\tilde{e}_M=\sum_{L=2}^{M} &(&-1)^{M-L}{N-L!\over N-M!\,M-L!}e_L R^{M-L}
\nonumber\\
+&(&-1)^{M-1}{N!(M-1)\over N-M!\,M!}R^M,
\end{eqnarray}
where we have noticed that, since $e_1=R$, the last two terms have the same
form and should be combined.
\hfil\break
\vskip 0.05truein
\noindent{\bf (2) Operating Hamiltonian $\mathbf\hat{H}$ on $\mathbf e_L$}
\vskip 0.05truein
\noindent
An important feature of $e_L$ is that it is also the monomial function
$m_{\lambda}$ corresponding to $\lambda=0^{N-L}1^L$. The normalised version
of $e_L$ can thus be written as $|\,e_L\rangle\equiv|\,0^{N-L}1^L\rangle$, 
and it is clear that the
Hamiltonian $\hat{H}$ can only connect this state to itself and
$|\,A\,\rangle\equiv|\,0^{N-L+1}1^{L-2}2^1\rangle$, where we have labelled
this state as $|\,A\,\rangle$ for convenience in what follows.
The two matrix elements can be calculated using from the formulas derived
in the previous sections for diagonal and off-diagonal elements respectively.
The diagonal element is given by
\begin{equation}
\langle e_L|\hat{H}|e_L\rangle=
=N^2-N-{1\over 2}L(L-1).
\end{equation}
The off-diagonal element is found by using the rule for the case where
only three separate angular momentum states change (here $1+1 \rightarrow 0+2$)
to give
\begin{equation}
\langle\,A\,|\,\hat{H}|\,e_L\rangle=
{1\over 2}\sqrt{2L(L-1)(N-L+1)}
\end{equation}
The final result for the operation of the Hamiltonian on $e_L$ is thus
\begin{eqnarray}
\label{Hamop}
\hat{H}|\,e_L\rangle&=&\left[N^2-N-{1\over 2}L(L-1)\right]
|\,e_L\rangle\nonumber\\
&+&{1\over 2}\sqrt{2L(L-1)(N-L+1)}|\,A\rangle.
\end{eqnarray}
\vskip 0.05truein
\noindent{\bf (3) Removing the Normalisation Factors}
\vskip 0.05truein
\noindent
We want to get rid of the normalisation factors for the eigenstates so
that the Hamiltonian will act directly on symmetric polynomials such
as $e_L$. In the previous equation we should divide by the normalization
factor for $|\,0^{N-L}1^L\rangle$ and multiply by that for 
$|\,0^{N-L+1}1^{L-2}2^1\rangle$. Using the form for the normalisation
factors given in Eq. (\ref{normwfn}), we find that the normalised version
of Eq. (\ref{Hamop}) is
\begin{equation}
\label{Hamopnorm}
\hat{H}|e_L)=[N^2-N-{1\over 2}L(L-1)]|e_L)
+{N-L+1\over 2}|A),
\end{equation}
where the $|\lambda)$ are the symmetric polynomials with no normalisation
factor.
\hfil\break
\vskip 0.05truein
\noindent{\bf (4) Relating $\mathbf|\,0^{N-L+1}1^{L-2}2^1)$ to $\mathbf e_L$
and $\mathbf Re_{L-1}$}
\vskip 0.05truein
\noindent
Consider the product
\begin{equation}
NR|e_{L-1})
=\left[\sum_{i=1}^N z_i\right]
\sum_{i_1<\dots<i_{L-1}}z_{i_1}z_{i_2}\dots z_{i_{L-1}}
\end{equation}
The above product can clearly only produce $|0^{N-L}1^L)$ or
$|0^{N-L+1}1^{L-2}2^1)$, depending upon whether the $z_j$ from the prefactor
is not or is included in the set $(z_{i_1},z_{i_2}\dots z_{i_{L-1}})$.
Moreover we see that each element of $|0^{N-L+1}1^{L-2}2^1)$ can only be
made in one way, so that
\begin{equation}
NRe_{L-1}=|0^{N-L+1}1^{L-2}2^1)+Ce_L.
\end{equation}
To find the coefficient $C$ we just count terms: $NRe_{L-1}$ has
$(N+1)!/(N-L+1)!\,(N-1)!$ terms, $|A)$ has
$N!/(N-L+1)!(L-2)!$ terms, and $e_L$ has $N!/(N-L)!L!$ terms.
This leads to the result $C=L$, and hence
\begin{equation}
|A)=NRe_{L-1}-Le_L.
\end{equation}
Combining this with Eq. (\ref{Hamopnorm}) gives
\begin{equation}
\label{Hamopel}
\hat{H}e_L=
\left[N^2-\left(1+{L\over 2}\right)N\right]e_L
+{N(N-L+1)\over 2}Re_{L-1}.
\end{equation}
\vskip 0.05truein
\noindent{\bf (5) Operating $\mathbf\hat{H}$ onto $\mathbf\tilde{e}_M$}
\vskip 0.05truein
\noindent
If we now operate $H$ onto $\tilde{e}_M$ using Eq. (\ref{emexpand}) to write
$\tilde{e}_M$ in terms of $e_L$ and $R$, and using Eq. (\ref{Hamopel})
to operate $\hat{H}$ onto $e_L$, we get after a lot of tedious algebra,
\begin{equation}
\label{result}
\hat{H}\tilde{e}_M=\left[N^2-\left(1+{M\over 2}\right)N\right]\tilde{e}_M.
\end{equation}
Note that we have to compare the coefficients of three types of term
separately: (i) $e_M$, (ii) $R^{M-L}e_L$ for $2\le L<M$, and (iii) $R^M$.
The proof can be simplified a little if we use the subspace property.
Introduce a projection operator, $\hat{P}$, which removes any term containing
a factor of $R$. Operating $\hat{P}$ onto Eq. (\ref{emexpand}) gives 
$\hat{P}e_L=\tilde{e}_L$, whilst operating $\hat{P}$ onto Eq. (\ref{Hamopel})
gives Eq. (\ref{result}).

\section{Discussion and Conclusions}

We have considered $N$ bosons in a 2D harmonic potential interacting via
repulsive delta-function potentials and with fixed total angular momentum
$L\le N$.
Within the ``lowest Landau level'' approximation, we have analytically
shown that the elementary symmetric polynomial of
coordinates relative to the center of mass, $\tilde{e}_L$ is an exact
eigenstate of this Hamiltonian with eigenvalue $N(N-L/2-1)$. Extensive 
numerical analysis shows that this state is actually the ground state. 
This is not surprising since in the special case $L=N$ this is just the
one-vortex state discussed in Ref. \cite{Wil98} which is expected to be
the ground state by analogy to superfluid $^4$He. One can also see that
there is a sense in which $\tilde{e}_L$ distributes the angular momentum
equally between particles subject to the subspace property.
We have attempted to prove analytically that $\tilde{e}_L$ is the ground state, 
but have so far failed.
The situation is much harder than in the proof we presented in Ref. \cite{Wil98}
to show that the eigenstate with the largest eigenvalue corresponds to
all angular momentum being in the center of mass motion. The main problem
is that the eigenvector for the smallest eigenvalue has components of both
signs to reduce its energy, and frustration results between any set of three
basis states that are connected by $\hat{H}$: the difficulty is essentially
that of a quantum antiferromagnet compared to a ferromagnet. For completeness
we hope that the state considered in this paper will be analytically shown to
be the ground state in the future, but from numerics there can be little
doubt that it is.
\vskip 0.15truein
\centerline{\bf ACKNOWLEDGEMENTS}
\vskip 0.15truein
We thank N.R. Cooper, J.M.F. Gunn and M.W. Long for useful discussions.
We acknowledge financial support from the EPSRC
Grants GR/M98975, GR/L28784 and the Nuffield Foundation.

\end{multicols}

\end{document}